# *In situ* GaN decomposition analysis by quadrupole mass spectrometry and reflection high-energy electron diffraction

S. Fernández-Garrido[*]

*ISOM and Dpt. de Ingeniería Electrónica, ETSI Telecomunicación*

*Universidad Politécnica de Madrid, 28040 Madrid, Spain*

G. Koblmüller

*Materials Department, University of California, Santa Barbara, CA 93106-5050, USA*

E. Calleja

*ISOM and Dpt. de Ingeniería Electrónica, ETSI Telecomunicación*

*Universidad Politécnica de Madrid, 28040 Madrid, Spain*

J. S. Speck

*Materials Department, University of California, Santa Barbara, CA 93106-5050, USA*

Thermal decomposition of wurtzite (0001)–oriented GaN was analyzed: in vacuum, under active N exposure, and during growth by rf-plasma assisted molecular beam epitaxy. The GaN decomposition rate was determined by measurements of the Ga desorption using *in situ* quadrupole mass spectrometry, which showed Arrhenius behavior with an apparent activation energy of 3.1 eV. Clear signatures of intensity oscillations during reflection high-energy electron diffraction measurements facilitated complementary evaluation of the decomposition rate and highlighted a layer-by-layer decomposition mode in vacuum. Exposure to active nitrogen, either under vacuum or during growth under N-rich growth conditions, strongly reduced the GaN losses due to GaN decomposition.

---

[*] electronic mail: sfernandez@die.upm.es

## I. INTRODUCTION

III-nitrides have been the subject of intense research since the early 1990's when the first blue light emitting diode[1] (LED) was developed. Nowadays, devices based on III-nitrides, such as visible and ultraviolet LEDs and laser diodes (LDs), or high electron mobility transistors (HEMTs) are commercially available. One of the main challenges to achieve these goals was the epitaxial growth of high quality material due to the lack of appropriate substrates. Rf plasma-assisted molecular beam epitaxy (PA-MBE) offers several advantages over other epitaxial growth techniques including *in situ* monitoring of the growth mode and rate by reflection high-energy electron diffraction (RHEED), a low impurity incorporation, and the realization of atomically sharp interfaces.

Though PA-MBE has reached its maturity for the growth of III-nitrides, a more complete analysis of the growth dynamics and thermal effects has yet to be accomplished. The growth of GaN by PA-MBE is a metastable process because GaN is thermodynamically unstable at pressures typical in the molecular beam regime ($<10^{-4}$ Torr).[2,3] To grow GaN, the *forward reaction* rate ($v_f$):

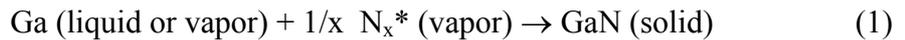

$$\text{Ga (liquid or vapor)} + 1/x\ N_x^* \text{(vapor)} \rightarrow \text{GaN (solid)} \qquad (1)$$

where $N_x^*$ represents any of the possible reactive nitrogen species, must be higher than the GaN decomposition rate or *reverse reaction ($v_r$)*:

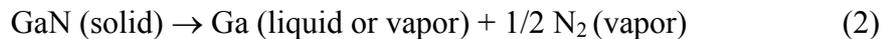

$$\text{GaN (solid)} \rightarrow \text{Ga (liquid or vapor)} + 1/2\ N_2 \text{(vapor)} \qquad (2)$$

The forward reaction rate is determined by the fluxes of the reactive species whose values are limited by the molecular beam regime, while the reverse reaction rate depends on the substrate temperature through the kinetic barrier for GaN decomposition.[2,3] Thus, for a given Ga/N flux ratio, the maximum growth temperature is given by $v_f - v_r$.

Therefore, one of the key issues toward the development of a high temperature growth diagram and toward improvements in the quality of the GaN grown by PA-MBE is to understand the mechanism behind GaN decomposition and its possible dependence on the growth parameters.

This work reports on the temperature dependence of wurtzite (0001)-oriented GaN decomposition (in the range from 720 to 805 ºC) under different conditions: i) in vacuum (at $N_2$ partial pressure of $1.5 \times 10^{-5}$ Torr); (ii) under active N exposure and iii) during standard GaN growth conditions[4] (Ga–rich and N–rich conditions). Two complementary techniques were employed to determine the GaN decomposition rate: line-of-sight quadrupole mass spectrometry (QMS) to measure the desorbed Ga flux[5,6] at elevated temperatures, and secondly RHEED intensity measurements to gain additional knowledge about the decomposition mode.

**II. EXPERIMENT**

The studies were carried out in a Varian Gen-II MBE system equipped with a Vecco Unibulb rf-plasma N source and two Ga Knudsen cells. During all experiments the $N_2$ partial pressure in the MBE system was $1.5 \times 10^{-5}$ Torr. As substrates we used 2–inch (0001) GaN templates (3.6 μm thick) grown by MOCVD on *c*–plane sapphire (Lumilog), which were outgassed for 1 hour at 400 ºC prior to the experiments. The substrate temperature was measured using an optical pyrometer calibrated to the melting point of Al (660 °C). Measurements of the dependence of the desorbing Ga flux on substrate temperature during GaN decomposition in vacuum, under active N exposure and during PA-MBE growth were performed using the QMS technique, which is described in detail elsewhere.[5,6] Both the QMS response flux and all other impinging fluxes are given in (0001) GaN growth units in *nm/min*.[4,6] We note that 1nm/min is equivalent to 0.064 ML/s, where one monolayer (ML) of GaN corresponds to c/2 =

0.259 nm or 1.14×10$^{15}$ GaN/cm$^2$ areal density along the (0001) direction. The RHEED pattern and intensity profile was monitored using a charge-coupled device (CCD) detection system (k-Space Associates). To ensure reproducible data, recovery of the GaN surface was achieved by growing a few MLs of GaN under Ga-rich conditions slightly below the limit of Ga droplets formation[4] prior to the initiation of each experiment.

## III. RESULTS

### A. GaN decomposition in vacuum

Since GaN decomposes congruently in vacuum,[7] the desorbing Ga flux measured by QMS ($\Phi_{QMS}$) provided a direct measurement of the GaN decomposition rate. Figure 1a. depicts the steady state desorbing Ga flux recorded at substrate temperatures intervals of ~ 20 °C between 720 and 805 °C. As shown in Fig. 1b., the GaN decomposition rate derived from the desorbing Ga flux increased exponentially with temperature from nearly zero at 720 °C to above ~ 3.5 nm/min (i.e. 0.2 μm/hr) at 805 °C. The onset temperature and the rates for thermal decomposition compare favourably with previous results from laser reflectivity.[8]

During GaN decomposition a streaky RHEED pattern was observed with clear intensity oscillations along the [11$\bar{2}$0] azimuth for temperatures above 740 °C (Fig. 2a). With increasing temperature the period of oscillations $\tau_{osc}$ became gradually shorter. Since each oscillation period corresponds to the decomposition of one ML of GaN, we were able to determine the decomposition rate for each temperature from the analysis of the RHEED intensity evolution. The decomposition rate given by $1/\tau_{osc}$, was nearly zero below 740 °C and increased exponentially with temperature exceeding a rate of ~ 3.5 nm/min at 805 °C (Fig. 2b.), in close agreement with the values determined by QMS. The observation of intensity oscillations can be further attributed to a two–dimensional (2D) layer–by–layer decomposition mode, similar to the typical growth rate oscillations during GaN growth.[9] In contrast, Grandjean et al.[8] also reported a streaky RHEED pattern during GaN decomposition but observed no intensity oscillations, suggesting rather a step-flow decomposition mode. We further note that the streaky RHEED pattern persisted well beyond the damping of the intensity oscillations (at least for others 60 seconds until the next surface recovery cycle), indicating the continuity of the 2D decomposition mode with no roughening of the GaN surface.

The Arrhenius plot of data points for the GaN decomposition rate (as evaluated by both QMS and RHEED) (Fig. 3) revealed a temperature dependence for the GaN decomposition rate, $\Phi_D$, defined as,

$$\Phi_D = A \cdot \exp(-E_D / k_B T) \qquad (3)$$

where the exponential prefactor, $A$, was fitted to $(1.6 \pm 0.1) \times 10^{15}$ nm/min and the apparent activation energy for GaN thermal decomposition was best fitted to $E_D = 3.1 \pm 0.1$ eV. This is in good agreement with values reported in the literature, ranging from the 3.1 to 3.6 eV.[7,8,10,11]

**B. GaN losses during PA-MBE growth and in vacuum under active N exposure**

The GaN losses as result of GaN decomposition under standard GaN growth conditions, as defined by the different growth conditions within the previously developed GaN growth diagram,[4] and in vacuum under active N exposure were also studied as a function on the substrate temperature. In particular, we investigated (i) intermediate Ga-rich conditions, where the GaN surface is terminated by a Ga adlayer [Ref. 4–6] ($\Phi_{Ga}$ = 7.0 nm/min, = 5.6 nm/min, = 5.0 nm/min for $\Phi_N$ = 5.0 nm/min and $\Phi_{Ga}$ = 10 nm/min, = 8.0 nm/min for $\Phi_N$ = 7.0 nm/min), (ii) N-rich conditions ($\Phi_{Ga}$ = 4.0 nm/min, 3.0 nm/min, = 1.0 nm/min for $\Phi_N$ = 5.0 nm/min and $\Phi_{Ga}$ = 6.0 nm/min, = 5.0 nm/min, = 4.0 nm/min for $\Phi_N$ = 7.0 nm/min) and (iii) exposure to active N without impinging Ga ($\Phi_{Ga}$ = 0, $\Phi_N$ = 5.0 nm/min, = 7 nm/min). In the following, to describe the actual GaN losses under these dynamics conditions, we introduce the term *effective GaN decomposition rate* defined as the flux of Ga or N atoms actually desorbed after GaN decomposition.

Under intermediate Ga-rich growth conditions a streaky RHEED pattern with intensity oscillations indicative of a 2D growth mode was observed for all temperatures, while under N-rich conditions we found a spotty–to–streaky transition with increasing growth temperature. This corresponds to a transition in growth mode from 3D to 2D layer–by–layer growth as recently highlighted by the sustainable 2D surface morphologies of layers grown under N–rich conditions in the GaN thermal decomposition regime [Ref. 12,13]. During GaN decomposition under active N exposure, the RHEED pattern became gradually spotty indicating progressive roughening of the GaN surface. This inhibited the assessment of the effective GaN decomposition rate under active N exposure using RHEED intensity analysis, in contrast to the layer–by–layer like decomposition observed in vacuum under static conditions (i.e. no impinging fluxes).

For those growth conditions where clear RHEED intensity oscillations were observed, we were able to investigate the effect of the Ga/N flux ratio on the effective GaN decomposition rate. Figure 4a. shows the actual growth rate given by $1/\tau_{osc}$ (where $\tau_{osc}$ is the RHEED intensity oscillation period), at 770 ºC as a function on the impinging Ga flux for two different active N fluxes ($\Phi_N$ = 5.0 and 7.0 nm/min). In both cases, the growth rate steadily increased until flux stoichiometry was achieved ($\Phi_{Ga} = \Phi_N$). An estimation of the effective GaN decomposition rate was made assuming a unity sticking coefficient for the impinging Ga and N atoms for N-rich and Ga-rich growth conditions respectively (Fig. 4b.). Thus, the effective GaN decomposition rate was evaluated as the nominal growth rate (given by the fixed impinging fluxes $\Phi_N$ or $\Phi_{Ga}$ for Ga-rich or N-rich growth conditions, respectively) minus the actual growth rate derived from the RHEED intensity oscillation periods (Fig. 4a.). For a given active N flux, the effective GaN decomposition rate steadily increased with the impinging Ga flux under N-rich conditions and saturated near 1 nm/min once flux stoichiometry was achieved. For a given Ga flux within the N-rich regime, the effective decomposition rate decreased with the active N flux. These results suggest a strong reduction of the effective GaN decomposition rate under N-rich conditions, towards higher N excess. Also, it is obvious that for slightly Ga–rich conditions the effective decomposition rate is independent of the Ga/N flux ratio (both by excess Ga and active N) and matches the GaN decomposition rate determined under static conditions (in vacuum, no growth). Nevertheless, we note that slight deviations between the actual effective GaN decomposition rates and those values shown in Fig. 4b. are expected due to the Ga sticking coefficient could be lower than unity for the studied range of temperatures and depend on the specific active N excess as further discussed in the following section.

To corroborate RHEED results, the effective GaN decomposition rate during PA-MBE growth and in vacuum under active N exposure was also investigated by QMS. Figure 5 shows the desorbed Ga flux, $\Phi_{QMS}$, under these dynamic conditions. As expected, in all cases $\Phi_{QMS}$ was found to increase exponentially with substrate temperature. A fitted curve for $\Phi_{QMS}$ during GaN decomposition in vacuum (i.e. $\Phi_D$, no impinging fluxes) as of Fig. 1b. and Fig. 2b. is also included for reference. Under slightly Ga-rich growth conditions, $\Phi_{QMS}$ was larger than $\Phi_D$, due to the typical excess

Ga desorbing from the surface,[4] as further discussed below. During N-rich growth and under active N exposure $\Phi_{QMS}$ was well below $\Phi_D$ and decreased with lower Ga/N flux ratios. These results point thus to a significant reduction of the effective GaN decomposition rate under active N excess (i.e. during N-rich growth or under active N exposure) in good agreement with the results derived from the analysis of the RHEED intensity evolution.

**IV. DISCUSSION**

To explain these deviations from the static conditions, we stress that under the dynamic conditions the desorbed Ga flux may not only arise from the Ga adatoms desorbing during GaN decomposition, but also from partial desorption of the impinging Ga flux during growth. The latter holds especially true for the excess Ga under Ga-rich conditions, but also for excess active N conditions, where the reduced Ga surface residence lifetime at elevated temperatures may limit the incorporation probability. For the Ga adatoms segregated to the surface under decomposition they can be either desorbed (as in the Ga–rich case) or re-incorporated into the layer (as in the N-rich case) via reaction with active N thus reducing the effective GaN decomposition rate.

Under slightly Ga-rich growth conditions the active N is assumed to be entirely consumed by the impinging Ga flux [Ref. 4]. Thus, under steady-state conditions, if there is no metal accumulation on the surface (as in form of Ga droplets) and the effective GaN decomposition rate is unaffected by the impinging fluxes, the desorbing Ga flux should be given by:

$$\Phi_{QMS} = \Phi_{Ga} - \Phi_N + \Phi_D \qquad (4)$$

meaning that $\Phi_{QMS}$ is given by the excess of the impinging Ga flux plus the GaN decomposition rate measured in vacuum. As evident in Fig. 5, this properly reproduces the experimental data without any fitting. Thus, we conclude that the effective GaN decomposition rate under slightly Ga–rich conditions is identical to the GaN decomposition rate found in vacuum (i.e. $\Phi_D$). Slightly Ga–rich conditions therefore result in unmodified GaN losses, as long as the GaN surface is not saturated with liquid Ga. The Ga coverage under these investigated temperatures and Ga fluxes was measured to be below 1 ML [Ref. 12], far away from the formation boundary of liquid Ga droplets. Liquid Ga droplets were considered to act as a catalyst for GaN decomposition[14] and were found to be correlated with the GaN decomposition enhancement observed in $H_2$ near atmospheric pressure.[15] Furthermore, given that the absolute value of the GaN decomposition rate approaches the N–limited GaN growth rate (i.e. 5.0 nm/min) at 810 °C, this temperature marks an upper limit for the successful growth of GaN for this fixed nitrogen flux and if exceeded results in thermal etching. It is therefore expected that higher active nitrogen fluxes will increase the Ga incorporation rate and thus enable growth at even higher temperatures.

In contrast, the results derived from both RHEED and QMS demonstrate that the effective GaN decomposition rate is strongly reduced under active N excess. From a thermodynamic point of view, the reduction of the effective GaN decomposition rate under such conditions can be explained considering Le Chatelier`s principle. For a given temperature and impinging Ga flux, an increase of the active N flux means a higher concentration of the reactives in equation (1). This yields a shift of the reaction in the opposite direction to counter-act the imposed change as long as there is Ga available. Thus, the ratio between the Ga adatoms which are reincorporated into the layer (after GaN decomposition) and those which are desorbed from the surface is increased for higher active N fluxes resulting in a reduction of the effective GaN decomposition rate. These results highlight the possibility to grow GaN under N-rich conditions at higher temperatures than expected from the GaN decomposition rate measured in vacuum, paving the way to an alternative approach to grow high quality material by PA-MBE.[13]

From a surface kinetic point of view, the differences between the GaN decomposition rate in vacuum (or under slightly Ga–rich growth conditions) and the effective decomposition rate under N–rich growth conditions can be understood by the capture probability and surface diffusion of the decomposed species with respect to the excess versus deficiency of N atoms (see Fig. 6). This is primarily based on the assumption that the decomposed species (i.e. those Ga and N atoms resulting from GaN decomposition) follow the mechanistic desorption pathway from solid to vapor via an adsorbed surface state:

$$\text{GaN (solid)} \rightarrow \text{Ga (adsorbed)} + \text{N (adsorbed)} \qquad (5)$$

$$\text{Ga (adsorbed)} \rightarrow \text{Ga (vapor)},$$

$$\text{N(adsorbed)} + \text{N(adsorbed)} \rightarrow \text{N}_2 \text{ (vapor)}$$

Proof for these intermediate adsorbed surface states of segregated Ga and N adatoms is directly provided by the reduction in the effective GaN decomposition rate due to their reincorporation under excess N conditions. If GaN decomposes directly from the solid to the vapor phase, the effective decomposition rate would not have been influenced at all by active nitrogen, yielding constant rates throughout all possible growth conditions. It is important to note, that the segregated Ga atoms from GaN decomposition cannot accumulate as liquid Ga (in the form of droplets), since the rate for thermal GaN decomposition is far below the desorption rate from a liquid Ga reservoir.[4,9]

In the simple case of decomposition into vacuum or under stoichiometric flux conditions, the rate of thermal GaN losses is therefore determined by the immediate evaporation of the segregated Ga and N adatoms due to the lack of excess Ga or N atoms provided to the surface.

However, under N–rich conditions the capture and reincorporation probabilities between segregated Ga and N adatoms are increased, partly due to the higher density of excess N and the modified surface diffusion behavior under these conditions. The latter can be understood by the fact, that N adatoms are expected to be kinetically stabilized on the surface, although they are thermodynamically highly unstable against desorption as $N_2$ molecules. According to Zywietz et al.[Ref. 16], the migration of N adatoms on a surface saturated with excess N conditions is a highly activated process (i.e. with a calculated diffusion barrier of 1.4 eV) [Ref. 16], such that the subsequent formation and evaporation rate of $N_2$ may become smaller than the capture rate by faster moving Ga adatoms. Consequently, the mean diffusion length of the Ga adatoms is limited by the capture with N adatoms, explaining the observed reduction in desorption of Ga adatoms.

Under N deficient (i.e. Ga–rich) conditions, the surface kinetics behave in the opposite way. First, no excess active N is available to be captured by the segregated Ga adatoms, since all impinging N atoms are entirely consumed by the supplied Ga flux [Ref. 4]. Also, on a Ga–rich GaN surface the diffusion barrier of N adatoms is reduced (i.e. 0.2 eV as in the extreme case of a liquid Ga bilayer terminating the surface under heavy Ga–rich conditions) [Ref. 17]. This results in higher migration of N adatoms and formation rates of volatile $N_2$, which in turn means much lower capture probability with Ga adatoms and enhanced effective GaN decomposition rates. This may also explain the recent findings, that liquid Ga (either in the form of a bilayer or droplets) can even catalyze GaN decomposition.[14]

Considering the similarity between the decomposition rates into vacuum (i.e. no growth) and the effective decomposition rates observed under slightly Ga–rich conditions, we suggest that in the present study surface diffusion plays an inferior role due to the low amounts of excess Ga. Recent studies showed that the Ga adlayer coverage involved under these slightly Ga–rich conditions at high temperature is much less than 1 ML [Ref. 12], which is expected to have not too significant effects on the N adatom diffusion [Ref. 18]. Increases in Ga flux and larger adlayer coverages (> 2 ML) would be required to obtain more substantial deviations between the decomposition rate into vacuum and the effective decomposition rate during Ga-rich growth.

## V. CONCLUSIONS

GaN decomposition was studied *in situ* by RHEED and QMS in vacuum with and without active N exposure, and during GaN PA-MBE growth (under slightly Ga-rich and N-rich growth conditions). RHEED intensity oscillations during GaN decomposition in vacuum revealed that GaN decomposes by a layer-by-layer mode. From both the period of RHEED intensity oscillations and the desorbing Ga flux assessed by QMS, the temperature dependent rates for GaN decomposition were determined and an apparent activation energy of 3.1 eV was deduced. Growth under slightly Ga-rich conditions yielded an effective GaN decomposition rate identical to the GaN decomposition rate measured in vacuum. In contrast, under N-rich growth conditions or exposing the GaN surface to active N, the effective GaN decomposition rate was substantially suppressed. This effect was found to be enhanced for higher active N excess due to the increase of Ga capture probability by N adatoms.

**ACKNOWLEDGEMENTS**

Thanks are due to A. Hirai, C. Gallinat, A. Corrion and C. Poblenz for fruitful discussions (all UCSB). The authors gratefully acknowledge support from AFOSR (Donald Silversmith, Program Manager). This work made use of the MRL Central

Facilities and was also partially supported by the Spanish Ministry of Education (MAT2004-2875, NAN04/09109/C04/2, Consolider-CSD 2006-19, and FPU program); and the Community of Madrid (GR/MAT/0042/2004 and S-0505/ESP-0200).

**References**

[1] S. Nakamura, and G. Fasol, The Blue Laser Diode-GaN Light emitters and Lasers, Springer, Berlin (1997).

[2] N. Newman, J. Ross, and M. Rubin, Appl. Phys. Lett. **62**, 1242 (1993).

[3] N. Newman, J. Cryst. Growth, **178**, 102 (1997).

[4] B. Heying, R. Averbeck, L. F. Chen, E. Haus, H. Riechert, and J. S. Speck, J. Appl. Phys. **88**, 1855 (2000).

[5] G. Koblmüller, R. Averbeck, H. Riechert, and P. Pongratz, Phys. Rev. B, **69**, 035325, (2004).

[6] J. S. Brown, G. Koblmüller, F. Wu, R. Averbeck, H. Riechert, and J. S. Speck, J. Appl. Phys. Lett. **99**, 074902 (2006).

[7] Z. A. Munir, and A. W. Searcy, J. Chem. Phys. **42**, 4223 (1965).

[8] N. Grandjean, J. Massies, F. Semond, S. Yu. Karpov, and R. A. Talalaev, Appl. Phys. Lett. **74**, 1854 (1999).

[9] C. Adelmann, J. Brault, G. Mula, B. Daudin, L. Lymperakis, and J. Neugebauer, Phys. Rev. B **67**, 165419 (2003).

[10] R. Groh, G. Gerey, L. Bartha, and J. I. Pankove, Phys. Status Solidi, **26**, 353 (1974).

[11] R. Held, D. E. Crawford, A. M. Johnston, A. M. Dabiran, and P. I. Cohen, Surf. Rev. Lett. **5**, 913 (1998).

[12] G.Koblmüller, S. Fernández-Garrido, E. Calleja, and J. S. Speck, Appl. Phys. Lett., **91**, 161904, (2007).

[13] G.Koblmüller, F. Wu, T. Mates, J. S. Speck, S. Fernández-Garrido, and E. Calleja, Appl. Phys. Lett., **91**, 221905, (2007).

[14] A. Pisch and R. Schmid –Fetzter, J. Cryst. Growth **187**, 329 (1998).

[15] D. D. Koleske, A. E. Wickenden, R. L. Henry, M. E. Twigg, J. C. Culbertson, and R.

J. Gorman, Appl. Phys. Lett. **73**, 2018 (1998).

[16]T. Zywietz, J. Neugebauer, and M. Scheffler, Appl. Phys. Lett. **73**, 487 (1998).

[17]J. Neugebauer, T. Zywietz, M. Scheffler, J. E. Northrup, H. Chen, and R. M. Feenstra, Phys. Rev. Lett. **90**, 056101 (2003).

[18]G.Koblmüller, J. Brown, R. Averbeck, H. Riechert, P. Pongratz, and J. S. Speck, Jpn. J. Appl. Phys. **44**, L906 (2005).

**List of Figures**

Figure 1  (a) Desorbing Ga flux assessed by QMS, $\Phi_{QMS}$, during GaN decomposition in vacuum at different substrate temperatures. (b) Dependence of the GaN decomposition rate on substrate temperature in vacuum as derived from $\Phi_{QMS}$. The solid line corresponds to the fit to equation (3).

Figure 2  (a) RHEED intensity oscillations during GaN decomposition in vacuum at different temperatures. (b) Dependence of the GaN decomposition rate on substrate temperature in vacuum as derived from the RHEED intensity oscillation periods, $\tau_{osc}$. The solid line corresponds to the fit to equation (3).

Figure 3  GaN decomposition rate in vacuum (as derived from both the desorbing Ga flux assessed by QMS and the RHEED intensity oscillation periods) as a function of $1/k_BT$. The activation energy for GaN decomposition is $E_D = 3.1 \pm 0.1$ eV.

Figure 4  GaN growth (a) and effective decomposition (b) rates at 770 ºC as a function on the impinging Ga flux for two different active N fluxes, 5.0 and 7.0 nm/min. The growth rate was assessed by RHEED intensity oscillations and the effective GaN decomposition rate was derived from the reduction of the nominal growth rate given by $\Phi_N$ or $\Phi_{Ga}$ for Ga-rich or N-rich growth conditions, respectively.

Figure 5   Dependence of the desorbing Ga flux ($\Phi_{QMS}$) on substrate temperature under slightly Ga-rich growth conditions ($\Phi_{Ga}$ = 7.0 nm/min, 5.6 nm/min, $\Phi_N$ = 5.0 nm/min), and under excess active N conditions ($\Phi_{Ga}$ = 3.0 nm/min, 1.0 nm/min, 0 nm/min; $\Phi_N$ = 5.0 nm/min). The black solid lines correspond to the plot of equation (4) for slightly Ga-rich conditions while the dashed lines are guides to the eye. The desorbing Ga flux during GaN decomposition in vacuum was also included for comparison (grey solid line).

Figure 6   Schematic of the GaN surface during GaN decomposition in vacuum (a), under slightly Ga-rich growth conditions (b), and under active N excess conditions (i.e. during N-rich growth or in vacuum under active N exposure) (c). The mechanistic desorption and reincorporation (for active N excess conditions) pathways for the decomposed species are shown in each case. The impinging fluxes in (b) and (c) are not depicted for clarity purposes.

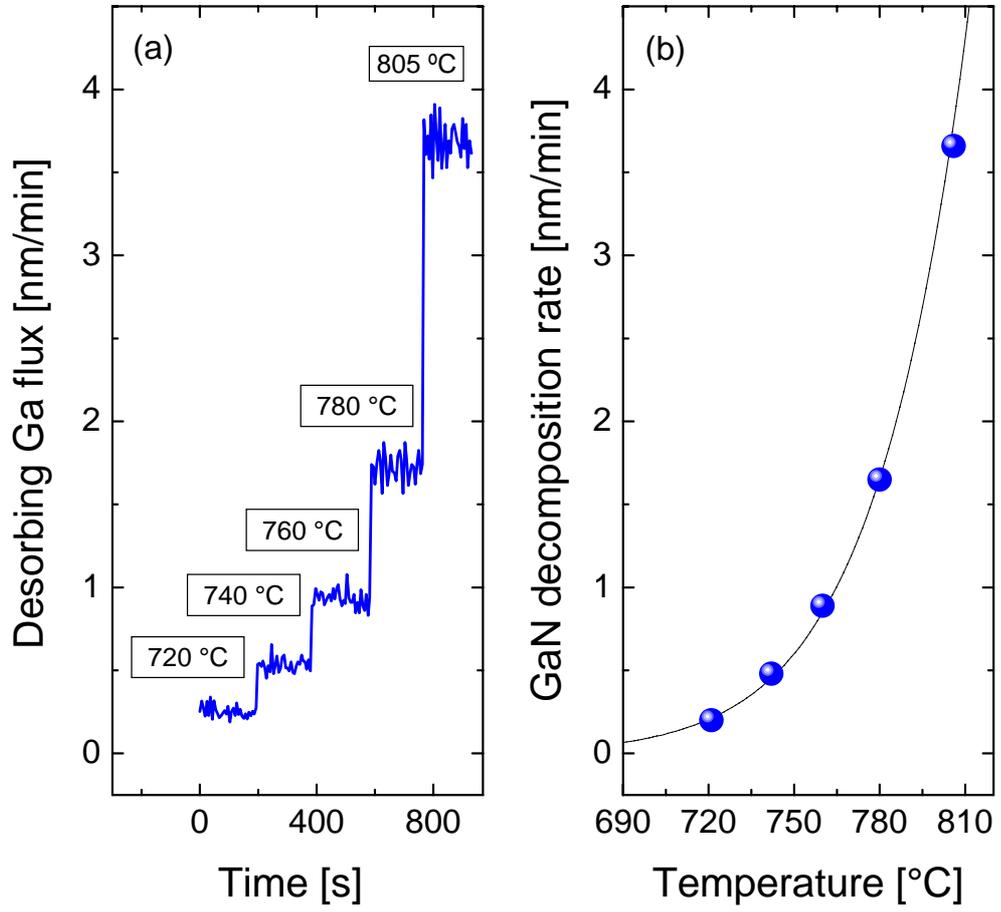

Figure 1

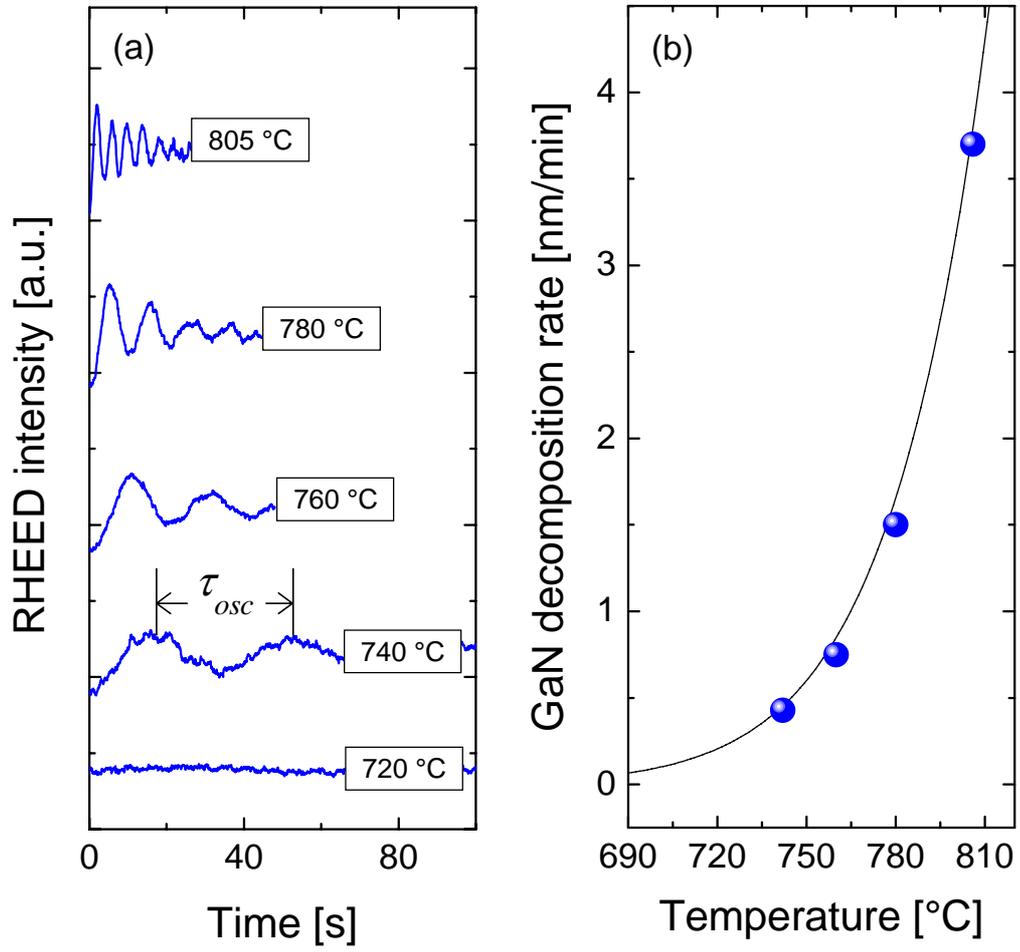

Figure 2

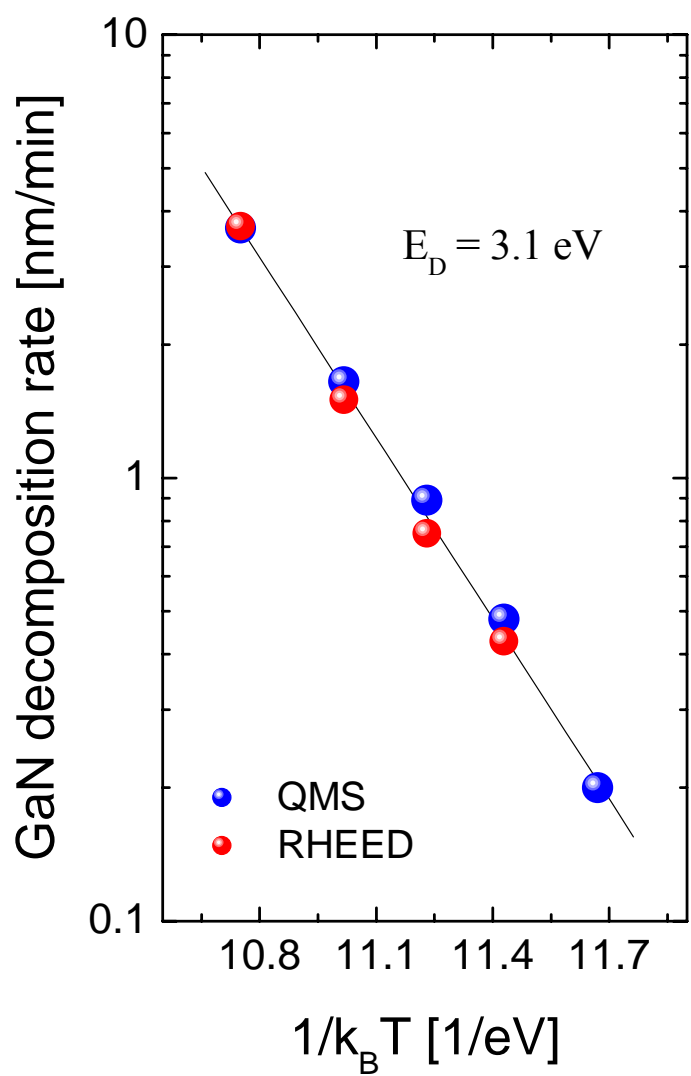

Figure 3

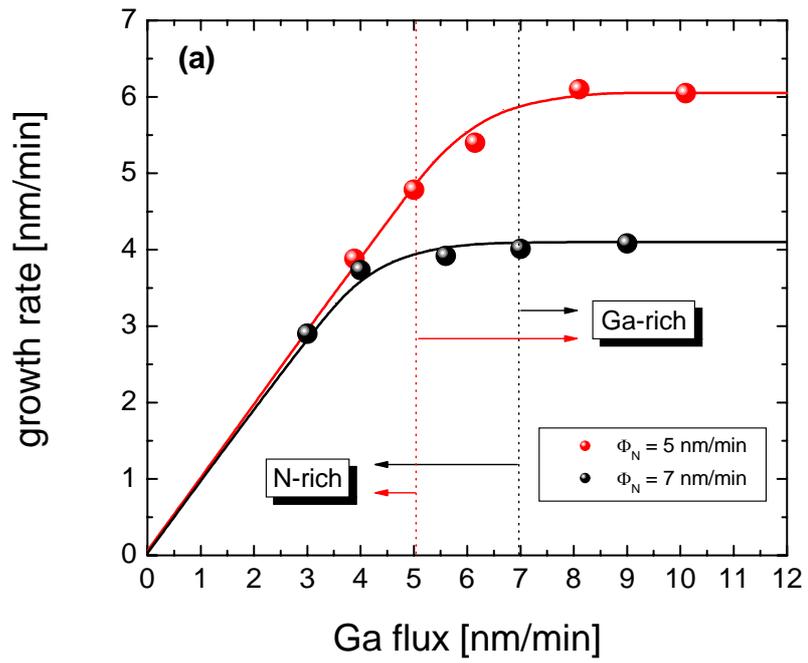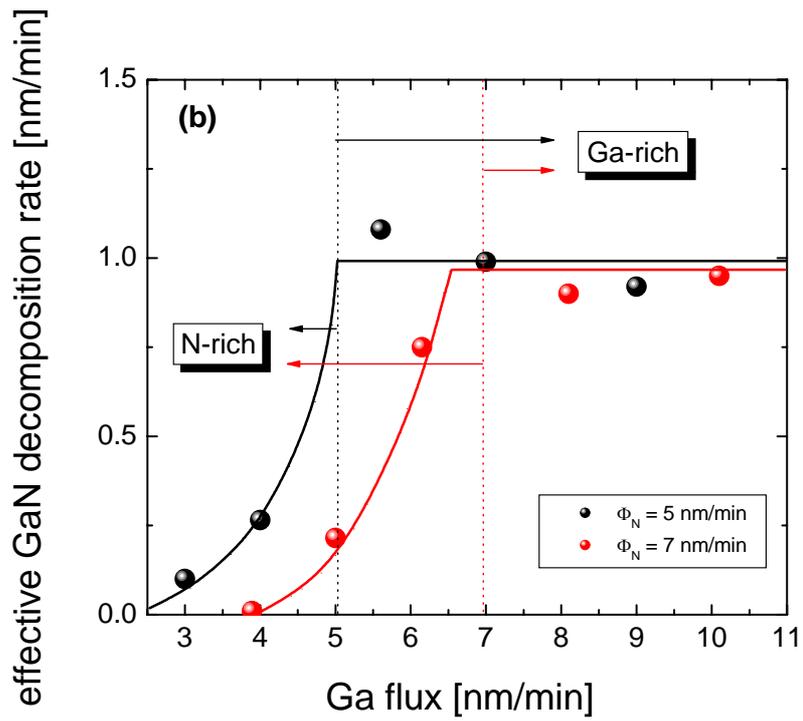

Figure 4

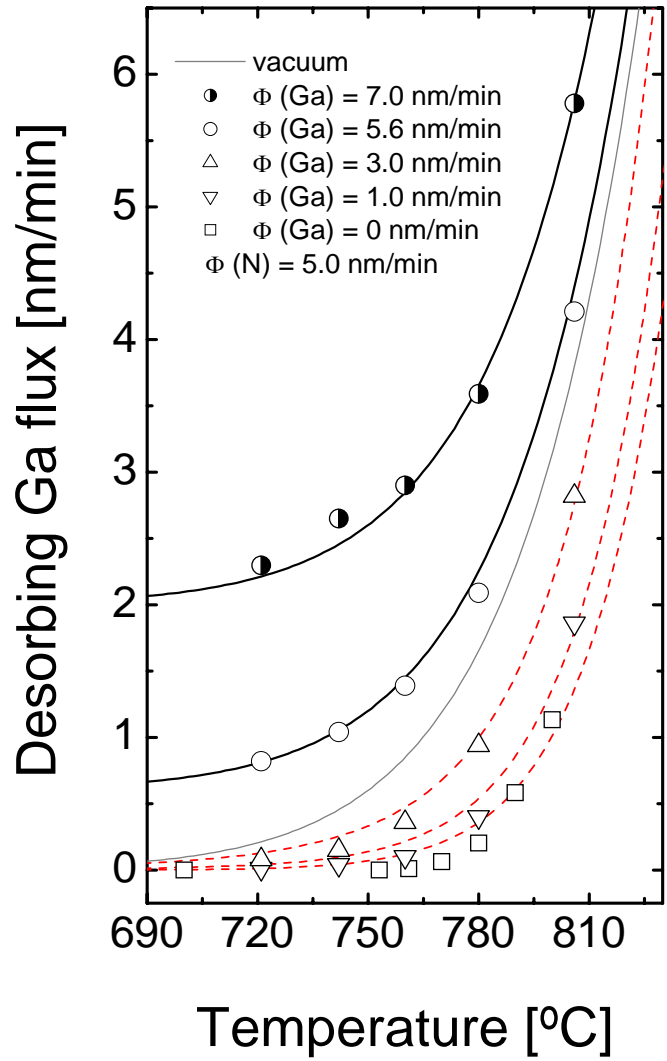

Figure 5

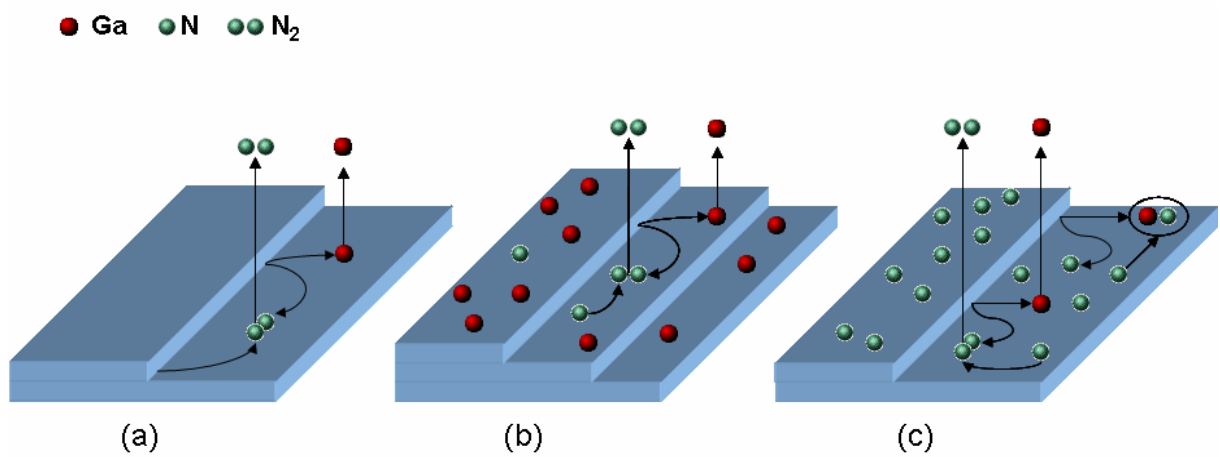

Figure 6